\documentclass[a4paper,11pt]{article}

\usepackage{jcappub, bm, color} 
\usepackage{amssymb,amsfonts,slashed,amsthm,amsmath,graphicx, soul}
\usepackage[caption=false]{subfig}
\usepackage{epsfig}

\def\beqa{\begin{eqnarray}}
\def\eeqa{\end{eqnarray}}
\def\p{\partial}

\def\p{\partial}

\newcommand{\lmk}{\left(}  
\newcommand{\rmk}{\right)}
\newcommand{\lkk}{\left[}  
\newcommand{\rkk}{\right]}

\newcommand{\bea}{\begin{array}}
\newcommand{\eea}{\end{array}}
\newcommand{\beq}{\begin{eqnarray}}
\newcommand{\eeq}{\end{eqnarray}}

\newcommand{\abs}[1]{\left\vert {#1} \right\vert}

\newcommand{\eq}[1]{Eq.~(\ref{#1})}

\begin{document}

\begin{flushright}
MIT-CTP/5161
\end{flushright}

\title{4-volume cutoff measure of the multiverse}

\author{Alexander Vilenkin$^1$ and Masaki Yamada$^2$
\\*[20pt]
$^1${\it \normalsize 
 Institute of Cosmology, Department of Physics and Astronomy,\\ 
Tufts University, Medford, MA 02155, USA} \\*[5pt]
$^2${\it \normalsize
Center for Theoretical Physics, Laboratory for Nuclear Science and Department of Physics, 
\\
Massachusetts Institute of Technology, Cambridge, MA 02139, USA}  \\*[50pt]
}
\emailAdd{vilenkin@cosmos.phy.tufts.edu}

\emailAdd{yamada02@mit.edu}

\abstract{
Predictions in an eternally inflating multiverse are meaningless unless we specify the probability measure. The scale-factor cutoff is perhaps the simplest and most successful measure which avoid catastrophic problems such as the youngness paradox, runaway problem, and Boltzmann brain problem, but it is not well defined in contracting regions with a negative cosmological constant. In this paper, we propose a new measure with properties similar to the scale-factor cutoff which is well defined everywhere. The measure is defined by a cutoff in the 4-volume spanned by infinitesimal comoving neighborhoods in a congruence of timelike geodesics. The probability distributions for the cosmological constant and for the curvature parameter in this measure are similar to those for the scale factor cutoff and are in a good agreement with observations.
}

\maketitle

\section{Introduction}

Observational predictions in multiverse models depend on one's choice of the probability measure. Different measure prescriptions can give vastly different answers.  This is the so-called measure problem of eternal inflation. 
Perhaps the simplest way to regulate the infinities of eternal inflation is to impose a cutoff on a hypersurface of constant global time.  One starts with a patch of a spacelike hypersurface $\Sigma$ somewhere in the inflating region of spacetime and follows its evolution along the congruence of geodesics orthogonal to $\Sigma$.  The cutoff is imposed at a hypersurface of constant time $t$ measured along the geodesics.  The resulting measure, however, depends on the choice of the time variable $t$.

An attractive choice is to use the proper time $\tau$ along the geodesics~\cite{Linde:1993xx,GarciaBellido:1993wn,Vilenkin:1994ua}.  One finds, however, that this proper time measure suffers from the youngness paradox, predicting that the universe should be much hotter than observed~\cite{Tegmark:2004qd}.  Another popular choice is the scale factor time, $t=\ln a$, where $a$ is the expansion factor along the geodesics~\cite{Linde:1993xx,GarciaBellido:1993wn,DeSimone:2008bq, Bousso:2008hz, DeSimone:2008if}.  The problem with this choice is that the scale factor evolution is not monotonic.  For example, in regions with a negative cosmological constant, $\Lambda<0$, expansion is followed by contraction, so $a$ starts to decrease along the geodesics.  
The scale factor measure then requires that the entire contracting region to the future of the turnaround point be included under the cutoff.  This gives a higher weight to regions of negative $\Lambda$, so the scale factor measure tends to predict that we should expect to measure $\Lambda<0$ (unless this is strongly suppressed by anthropic factors). 
Some other measure proposals have even more severe problems with negative $\Lambda$.  For example, the lightcone time cutoff \cite{Bousso:2009dm} gives an overwhelming preference for $\Lambda < 0$ \cite{Salem:2009eh}.\footnote{Local measure proposals, which sample spacetime regions around individual geodesics with subsequent averaging over an ensemble of geodesics, yield probability distributions that sensitively depend on the choice of the ensemble.  This choice is largely arbitrary, and thus these proposals are incomplete as they now stand.  The ``watcher measure" of Ref.~\cite{Garriga:2012bc} follows a single ``eternal" geodesic, but makes the assumption that the big crunch singularities in AdS bubbles lead to bounces, where contraction is followed by expansion, so that geodesics can be continued through the crunch regions.  We do not adopt this assumption in the present paper.
}

In this paper, we introduce a new global time measure which does not suffer from these problems.
We divide the initial hypersurface $\Sigma$ into infinitesimally small segments of equal 3-volume $\epsilon \to 0$ and follow the evolution of these segments along the orthogonal congruence of geodesics.  The time coordinate $\Omega$ is defined as the 4-volume spanned by the segment,
\beq
\Omega(\tau) = \frac{1}{\epsilon} \int_{(0,\tau) \times \epsilon {\cal V}^{(3)} (\tau)} \sqrt{-g} \, d^4 x = \int_0^\tau d\tau' {\cal V}^{(3)}(\tau'),
\label{Omega0}
\eeq
where $\epsilon {\cal V}^{(3)}(\tau)$ is the 3-volume of the evolved segment at proper time $\tau$, $\tau$ is set equal to zero at $\Sigma$, and ${\cal V}^{(3)}(0)=1$.  $\Omega$ has a clear geometric meaning and it clearly grows monotonically along the geodesics.  The measure is defined by imposing a cutoff at $\Omega_c={\rm const}$.
If the universe can locally be approximated as homogeneous and isotropic, we can write ${\cal V}^{(3)}(\tau) = a^3(\tau)$, where $a(\tau)$ is the scale factor with $a(0)=1$.  Then
\beq
\Omega(\tau) = \int_0^\tau d\tau' a^3(\tau').
\label{Omega}
\eeq
We can think of the geodesics in the congruence as representing an ensemble of inertial observers spread uniformly over the initial surface $\Sigma$.  The measure prescription is then that each observer samples an equal 4-volume $\propto\Omega_c$.  

The distribution of ``observers" may become rather irregular in regions of structure formation. The scale factor (or the 3-volume ${\cal V}^{(3)}$ in Eq.~(\ref{Omega0})) comes to a halt in collapsed regions which have decoupled from the Hubble flow and continues to evolve between these regions.  
Furthermore, the geodesic congruence may develop caustics where geodesics cross.  One can adopt the rule that geodesics are terminated as they cross at a caustic.  As it was noted in Ref.~\cite{Bousso:2012tp}, this does not create any gaps in the congruence.  But the resulting cutoff surface would still be rather irregular.  Such dependence of the measure on details of structure formation appears unsatisfactory and calls for some sort of coarse graining, with averaging over the characteristic length scale of structure formation.  This issue was emphasized in Ref.\cite{Bousso:2008hz} in the case of scale factor measure and was further discussed in Ref.~\cite{DeSimone:2008if}.

A somewhat related problem is that even though $\Omega$ grows monotonically along geodesics of the congruence, the surfaces of constant $\Omega$ are not necessarily spacelike, so $\Omega$ is not a good global time coordinate.  As a result an event may be included under the cutoff, while some events in its causal past are not included.  A possible way to cure this problem is to modify the cutoff surface $\Omega=\Omega_c$ by excluding future light cones of all points on that surface.\footnote{This prescription was suggested in Ref.~\cite{DeSimone:2008if} to address a similar problem for the scale factor measure.}  Then all events under the cutoff are included together with their causal past.  This prescription also alleviates the problem of sensitivity of the measure to structure formation.  If the characteristic scale of structure formation is much smaller than the horizon, the modified cutoff surface would roughly coincide with a constant $\Omega$ surface in the background FRW geometry. 

The implementation of the 4-volume measure is somewhat more complicated than in the cases of proper time and scale factor measures, but it becomes tractable in a number of interesting special cases.  In the next section we use this measure to estimate the volume fraction occupied by different vacua in the eternally inflating part of spacetime, assuming low transition rates between the vacua. 
In Sections 3 and 4 we find respectively the probability distributions for the cosmological constant and for the density parameter (or spatial curvature) under assumptions similar to those that were used in Refs.~\cite{DeSimone:2008bq,DeSimone:2009dq} to calculate these distributions in the scale factor measure.  A formalism that can be used to determine the distributions in more general landscapes is outlined in Section 5.  Finally, our results are briefly summarized and discussed in Section 6.

\section{Volume distribution of vacua}

Consider a multiverse consisting of bubbles of de Sitter (dS) and terminal (Anti-de Sitter and Minkowski) vacua, labeled by index $j$. The expansion rate of dS vacuum $j$ is $H_j$ and nucleation rate of bubbles of vacuum $i$ in parent vacuum $j$ per Hubble volume per Hubble time is $\kappa_{ij}$.  We shall assume that $\kappa_{ij}\ll 1$ -- which is expected, since nucleation occurs by quantum tunneling.  In this section we shall calculate the 3-volume occupied by each dS vacuum on a surface of constant $\Omega$ in the inflating part of spacetime and use the result to find the abundances of Boltzmann brains in dS vacua.  
We shall not be interested in volumes occupied by terminal vacua in this section.

\subsection{Relation to scale factor cutoff}

An approximate relation between the 4-volume and scale factor cutoffs can be found if we note that the scale factor grows exponentially in the inflating regions, and therefore the integral in Eq.~(\ref{Omega}) is dominated by the upper limit.  In a region occupied by vacuum $j$, the scale factor is $a_j(\tau)=Ce^{H_j \tau}$ with $C={\rm const}$, 
so we can write approximately 
\beq
\Omega_j (\tau)\approx  \int^\tau a_j^3(\tau')d\tau' \approx  \frac{a_j^3(\tau)}{3H_j}.
\label{Omegaa}
\eeq
The cutoff surface at $\Omega=\Omega_c = {\rm const}$ can then be approximated as
\beq
\frac{a^3(\tau)}{3H_j} = \Omega_c,
\label{cutoff}
\eeq
so the 4-volume cutoff at $\Omega=\Omega_c$ is approximately equivalent to the scale factor cutoff at
\beq
t_c =  \frac{1}{3}\ln (3 H_j \Omega_c),
\label{relation}
\eeq
where the scale factor time is defined as $t=\ln a$.

The approximations (\ref{Omegaa}), (\ref{relation}) are accurate, as long as the cutoff surface does not pass within a few Hubble times of a transition from one vacuum to another (on the daughter vacuum side). The correction to Eq.~(\ref{Omegaa}) is $\sim a_i^3 /3H_i$, where $a_i$ is the scale factor at the time when the vacuum region $j$ being considered was created from a parent vacuum $i$.  If $H_j\lesssim H_i$, which is usually the case, this correction is negligible already at one Hubble time after the transition $i\to j$, when $a/a_i=e$ and the correction is $\lesssim e^{-3} \approx 1/20$.  The correction is more significant for large upward jumps with $H_j\gg H_i$.  In this case the condition for Eq.~(\ref{relation}) to be accurate is $a/a_i \gtrsim (H_j/H_i)^{1/3}\gg 1$.  This would happen on some segment of the cutoff surface if it lies within a scale factor time $t_{ji}\sim (1/3)\ln(H_j/H_i)$ of the transition from $i$ to $j$ (on the side of $j$).  We expect such segments to be rare -- both because large upward jumps are strongly suppressed and because the interval $t_{ji}$ is much shorter than the  scale factor time that geodesics typically spend in vacuum $j$.
Thus we expect the approximations (\ref{Omegaa}), (\ref{relation}) to hold for a generic cutoff surface.

Similar approximations should apply in spacetime regions where the Hubble parameter $H$ is not constant, but varies on a timescale much longer that $H^{-1}$ (e.g., in quantum diffusion or slow-roll regions).  In this case Eq.~(\ref{Omegaa}) is replaced by
\beq
\Omega (\tau) \approx  \frac{a^3(\tau)}{3H(\tau)}.
\label{Omegaa2}
\eeq

\subsection{Volume distribution and Boltzmann brains}

We can now find the volume distribution of different vacua.  We start with the volume distribution on constant scale factor surfaces and then rewrite the result on a constant 4-volume surface by using \eq{relation}.  The former distribution can be found from the rate equation (see, e.g., \cite{DeSimone:2008if})
\beq
\frac{dV_i}{dt}= 3 V_i + \sum_j M_{ij}V_j ,
\eeq
where $V_i(t)$ is the volume occupied by vacuum $i$ on a constant scale factor surface $t={\rm const}$ within a region of a fixed comoving size, $t$ is the scale factor time, 
\beq
M_{ij}=\kappa_{ij} -\delta_{ij}\kappa_{i}
\label{Mij}
\eeq
is the transition matrix, and 
\beq
\kappa_i=\sum_r \kappa_{ri}
\eeq
is the total decay rate of vacuum $i$ per Hubble volume per Hubble time.  
The late-time asymptotic solution of this equation for dS vacua $i$ is
\beq
V_i(t) = s_i e^{(3-q)t},
\label{Vj}
\eeq
where $q>0$ is the smallest solution of the eigenvalue equation
\beq
(\kappa_i-q)s_i=\sum_j \kappa_{ij} s_j 
\eeq
and $s_i$ is the corresponding eigenvector.

Substituting \eq{relation} in Eq.~(\ref{Vj}) we find
\beq
V_i (\Omega_c) = s_i (3 H_i \Omega_c)^{1-q/3}. 
\label{VolumeDistribution}
\eeq
$q$ is an exponentially small number, so to a good approximation we can write
\beq
V_i (\Omega_c) \propto s_i H_i . 
\label{VolumeDistribution1}
\eeq
This is the (approximate) asymptotic volume distribution in the 4-volume cutoff measure.
Compared to the scale factor measure, the volume of faster expanding vacua is enhanced by a factor $H_i$.

The distribution (\ref{VolumeDistribution1}) can be used to find the abundance of Boltzmann brains (BBs) in different dS vacua.  Suppose BBs are produced in vacuum $i$ at a rate $\Gamma_i^{BB}$ per unit spacetime volume.  The number of BBs $N_i^{BB}$ is then proportional to the total 4-volume in that vacuum.  With a scale factor cutoff at $t=t_c$ this volume is
\beq
V_i^{(4)}(t_c)=\int^{t_c}V_i(t)d\tau=H_i^{-1}\int^{t_c}V_i(t)dt =\frac{1}{3-q}H_i^{-1} s_i e^{(3-q)t_c}, 
\label{V4tc}
\eeq
where we have used Eq.~(\ref{Vj}).  Now, using Eq.~(\ref{relation}) to express $t_c$ in terms of $\Omega_c$, we find
\beq
V_i^{(4)}(\Omega_c)\propto s_i H_i^{-q}
\eeq
and
\beq
N_i^{BB}\propto \Gamma^{BB} s_i,
\eeq
where we have approximated $H_i^{-q}\approx 1$. 

The difference from the scale-factor cutoff measure, which gives \cite{DeSimone:2008if,Bousso:2008hz} $N_i^{BB}\propto \Gamma_i^{BB} H_i^{-1} s_i$ is only by a factor of $H_i$, which is not exponentially large. 
Thus the analysis of the Boltzmann-brane problem 
in the 4-volume cutoff measure is (almost) the same as that in the scale-factor measure. 
Since the problem can be evaded in the latter measure~\cite{Bousso:2008hz, DeSimone:2008if}, 
we conclude that the 4-volume cutoff measure may also be free from the Boltzmann-brane problem, depending on the properties of the landscape.  We expect the conditions for avoidance of the BB problem to be very similar to those in the scale factor measure.

\section{Probability distribution for cosmological constant}

In this section we calculate the probability distribution for the cosmological constant $\Lambda$ under the same assumptions that were used in Ref.~\cite{DeSimone:2008bq} for the scale factor measure. Specifically, we focus on a subset of bubbles that have (nearly) the same physical properties as our bubble, apart from the value of $\Lambda$.  We shall assume that the number of such bubble types in the landscape is very large, so the distribution of $\Lambda$ is nearly continuous.  After nucleation each bubble goes through a period of slow-roll inflation, followed by periods of radiation and matter domination, until $\Lambda$ eventually starts to dominate.  We will be interested in the values of $\Lambda$ for which this happens late in the matter era.

Let ${\tilde a}_\Lambda(\tau)$ be the scale factor in a region with a given value of $\Lambda$, where the proper time $\tau$ is measured from the moment of thermalization (end of inflation) and ${\tilde a}$ is normalized so that ${\tilde a}(0)=1$.  We can define a reference time $\tau_m$ such that $\tau_{eq}\ll\tau_m\ll\tau_\Lambda$, where $\tau_{eq}$ is the time of equal matter and radiation densities and $\tau_\Lambda$ is the time of $\Lambda$ domination.  Then the evolution before $\tau_m$ is the same in all regions, while after $\tau_m$ the scale factor is given by
\beq
{\tilde a}_\Lambda(\tau)= 
\left\{ 
\bea{ll}
\displaystyle{
{\tilde a}_m \lmk \frac{3}{2} H_\Lambda \tau_m \rmk^{-2/3}\sinh^{2/3}\left(\frac{3}{2} H_\Lambda\tau\right)
}&~~~~{\rm for}~~ \Lambda > 0
\vspace{0.3cm}
\\
\displaystyle{
{\tilde a}_m \lmk \frac{3}{2} H_\Lambda\tau_m \rmk^{-2/3}\sin^{2/3}\left(\frac{3}{2} H_\Lambda\tau\right)
}&~~~~{\rm for}~~ \Lambda < 0, 
\eea
\right.
\label{a1}
\eeq
where $H_\Lambda=\sqrt{|\Lambda|/3}$.  Here, ${\tilde a}_m = {\tilde a}(\tau_m)$; it depends on the evolution prior to $\tau_m$, but the quantity ${\tilde a}_m \tau_m^{-2/3}$ is independent of $\tau_m$ (and of $\Lambda$). 
A cutoff at $\Omega=\Omega_c$ in a bubble thermalized at $\Omega_*$ with a scale factor $a_*$ corresponds to a cutoff at proper time $\tau_c$, which can be found from 
\beq
\Omega_c=\Omega_*+ a_*^3 \int_0^{\tau_c} {\tilde a}_\Lambda^3(\tau)d\tau. 
\label{Omegac}
\eeq
From Eq.~(\ref{Omegaa2}) we can write
\beq
\Omega_*\approx \frac{1}{3H_*} a_*^3,
\eeq
where $H_*$ is the expansion rate at the end of slow-roll inflation in the bubble.  
Hence we can rewrite Eq.~(\ref{Omegac}) as
\beq
\Omega_c\approx \Omega_*\left[1+ 3H_* \int_0^{\tau_c} {\tilde a}_\Lambda^3(\tau)d\tau\right]. 
\label{Omegac2}
\eeq

The rest of the analysis closely follows Ref.~\cite{DeSimone:2008bq}, where references to earlier literature can also be found. 
The physical volume thermalizing in a scale factor time interval $dt_*$ in the spacetime region defined by the geodesic congruence is 
\beq
d{\cal V}_*\propto e^{\gamma t_*}dt_* ,
\eeq
where $t_*=\ln a_*$ and $\gamma=3-q\approx 3$.  Expressing $t$ in terms of $\Omega$, we have
\beq
d{\cal V}_*\propto \Omega_*^{\gamma-3} d\Omega_* \approx d\Omega_*,
\eeq
which says that thermalized volume is produced at approximately constant rate per unit 4-volume.

After thermalization, density perturbations grow, some fraction of matter clusters into galaxies, and observers evolve in some of these galaxies.
The probability distribution for $\Lambda$ is proportional to the number of observers 
in regions with that value of $\Lambda$. 
We assume that the number of observers is proportional to the number of large galaxies with mass $M \gtrsim M_G$ ($\sim 10^{12} M_\odot$).  Then the probability distribution can be expressed as 
\beq
P(\Lambda)\propto \int_0^{\Omega_c} F(\tau_c-\Delta\tau) {d\Omega_*},
\label{P}
\eeq
where $F(\tau)$ is the fraction of matter that clusters into large galaxies at proper time $\tau$ after thermalization, 
$\Delta\tau$ is the time required for observers to evolve, and $\tau_c$ is expressed in terms of $\Omega_c/\Omega_*$ from Eq.~(\ref{Omegac2}).  Introducing a new variable $X=\Omega_*/\Omega_c$, we can write
\beq
P(\Lambda) = N \int_0^{1} F(\tau_c(X)-\Delta\tau) {dX}, 
\label{P2}
\eeq
where $N$ is a normalization constant determined by $\int P(\Lambda) d \Lambda / \Lambda_{\rm obs} = 1$ with $\Lambda_{\rm obs}$ being the observed value of cosmological constant. 

In Eqs.~(\ref{P}) and (\ref{P2}), we implicitly assumed that $\Lambda >0$.  
When the landscape includes AdS vacua with $\Lambda <0$, some of the AdS regions will crunch prior to the cutoff, and such regions should be treated separately.  
The probability distribution for $\Lambda <0$ should be calculated from
\beq
 P (\Lambda) 
 &=& N \lkk 
 \int_0^{X_{\rm crunch}} F(\tau_c(X_{\rm crunch})-\Delta\tau) dX 
 + \int_{X_{\rm crunch}}^1 F(\tau_c(X)-\Delta\tau) dX
 \rkk 
 \\
 &=&
 N \lkk 
 X_{\rm crunch} F(\tau_{\rm crunch} -\Delta\tau) 
 + \int_{X_{\rm crunch}}^1 F(\tau_c(X)-\Delta\tau) dX
 \rkk, 
 \label{P3}
\eeq
where $X_{\rm crunch} \equiv X(\tau_{\rm crunch})$ and $\tau_{\rm crunch} \equiv 2 \pi / 3 H_\Lambda$. 

We will be interested in regions where $\tau_c\gg\tau_m$; then the integral in (\ref{Omegac2}) is dominated by the range $\tau\gg\tau_m$, so we can use Eq.~(\ref{a1}) for ${\tilde a}_\Lambda(\tau)$.  
This gives 
\beq
X^{-1}\approx 
\left\{ 
\bea{ll}
\displaystyle{
\frac{2H_i {\tilde a}_m^3}{9H_\Lambda^3 \tau_m^2} \left[\sinh(3H_\Lambda\tau_c)-3H_\Lambda\tau_c\right] 
}&~~~~{\rm for}~~ \Lambda > 0
\vspace{0.3cm}
\\
\displaystyle{
\frac{2H_i {\tilde a}_m^3}{9H_\Lambda^3 \tau_m^2} \left[-\sin(3H_\Lambda\tau_c)+3H_\Lambda\tau_c\right] 
}&~~~~{\rm for}~~ \Lambda < 0. 
\eea
\right.
\label{X1}
\eeq
Note that $\tau_c$ is assumed to be smaller than $\tau_{\rm crunch} \equiv 2 \pi / 3 H_\Lambda$ for $\Lambda < 0$.

\begin{figure}
\centering
  \includegraphics[width=0.45\linewidth]{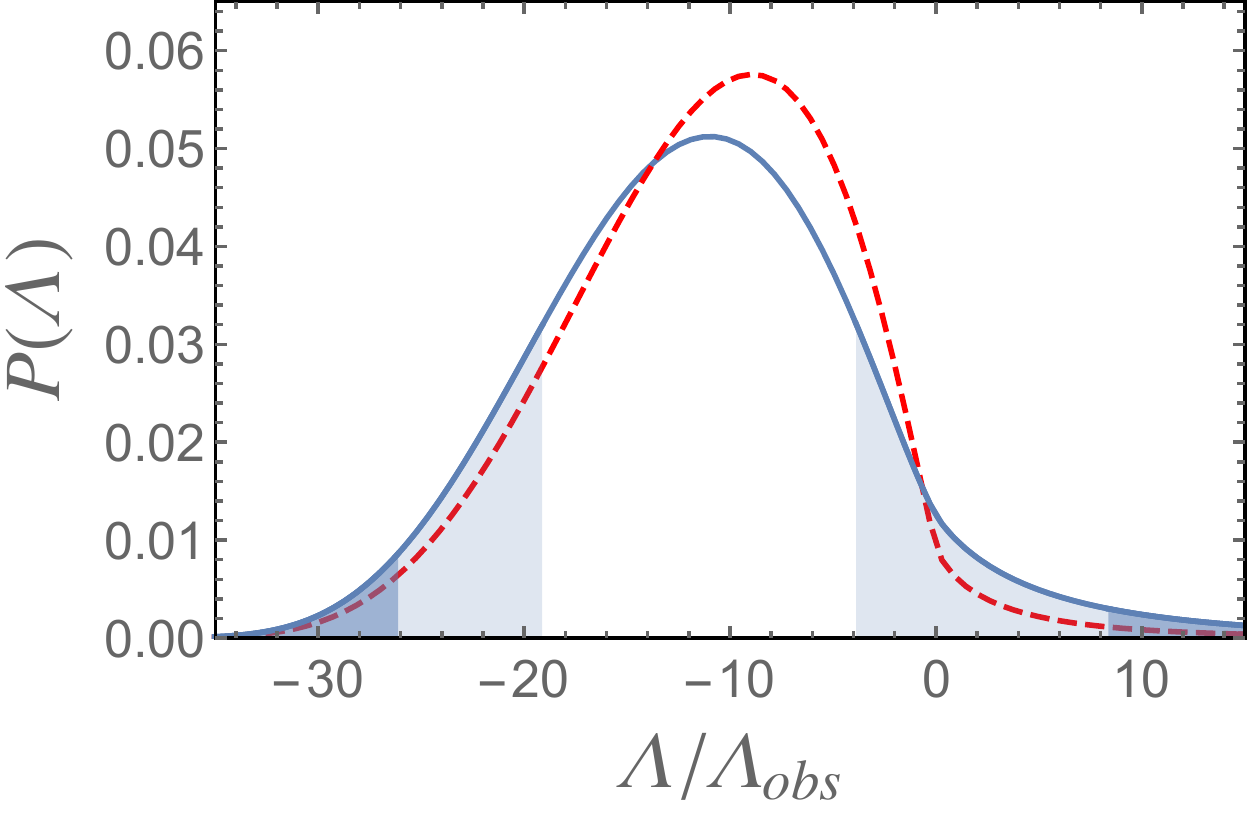}
  \qquad
  \includegraphics[width=0.45\linewidth]{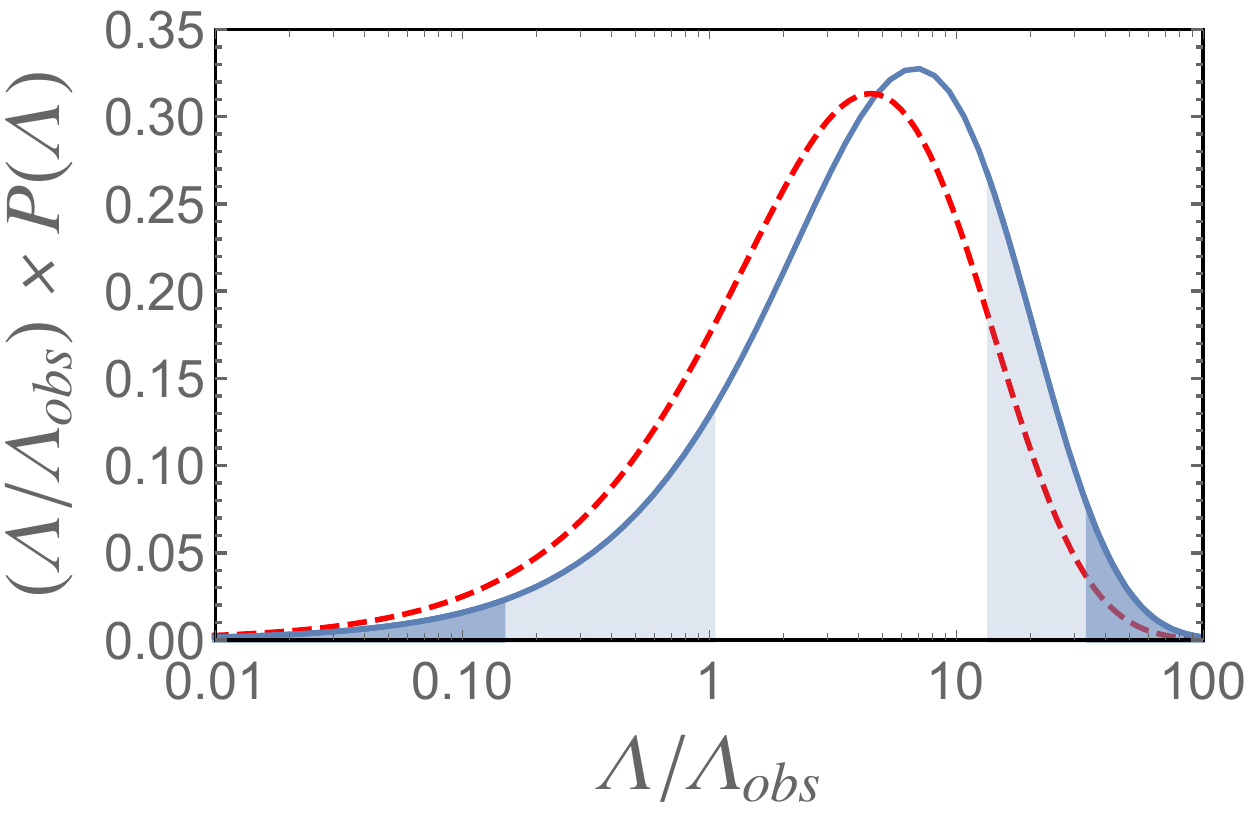}
 \caption{Distribution of cosmological constant in the 4-volume cutoff measure (solid blue curve) and the scale-factor cutoff measure (red dashed curve). 
The right panel is the probability distribution $\Lambda \times P(\Lambda)$ for $\Lambda>0$ in the logarithmic scale.  All distributions are normalized as $\int P(\Lambda)d\Lambda / \Lambda_{\rm obs}=1$. 
The lighter (darker) blue-shaded regions 
represent the $1\sigma$ ($2\sigma$) ranges for the probability distribution in the 4-volume cutoff measure. 
}
\label{fig1}
\end{figure}

We use the Press-Schechter form \cite{Press:1973iz,Bardeen:1985tr} with a linear perturbation theory 
for the collapsed fraction $F(\tau)$.  The distribution $P(\Lambda)$ can then be found numerically from Eqs.~(\ref{P2}) and (\ref{X1}), as it was done in Ref.~\cite{DeSimone:2008bq}.  
We use the same parameters as the one used in the same paper (e.g., $\Delta \tau = 5 \times 10^9$ years and the root-mean square fractional density contrast averaged over a comoving scale enclosing mass $10^{12} M_\odot$ at present $\sigma (10^{12} M_\odot) \approx 2.03$) while we use the updated cosmological parameters from the Planck data, such as $\Omega_\Lambda^{(\rm obs)} = 0.69$ and $\Omega_m^{(\rm obs)} = 0.31$~\cite{Aghanim:2018eyx}. 
We plot the resulting probability distributions in Fig.~\ref{fig1}, with solid blue and dashed red curves corresponding to 4-volume and scale factor cutoffs respectively.  The left panel shows the full distributions, while the right panel shows the (normalized) distributions for positive $\Lambda$ in the logarithmic scale.  
The lighter (darker) blue-shaded regions 
represent the $1\sigma$ ($2\sigma$) ranges for the probability distribution in the 4-volume cutoff measure.

To plot the distribution for $\Lambda<0$ in the scale factor measure, we set $\tau_c = \tau_{\rm crunch}$ for $\tau_c > \tau_{\rm turn}$, where $\tau_{\rm turn} \equiv \pi / 3 H_\Lambda$ is the turnaround time when the contracting phase begins and $\tau_{\rm crunch} \equiv 2\pi / 3 H_\Lambda$ is the time of the big crunch.  
Since $\tau_{\rm crunch}$ is twice larger than $\tau_{\rm turn}$, this results in a discontinuous jump of $\tau_c$ and in a larger probability for $\Lambda < 0$ in the scale-factor cutoff measure.  We see however that the difference between the distributions in the two measures is not dramatic.
The total probability for $\Lambda$ to be positive is $3\%$ for the scale factor and $8\%$ for the 4-volume cutoff measure.

We note that in either measure the probability of negative $\Lambda$ is expected to be significantly reduced due to anthropic effects that have not been taken into account here.  After the turnaround galaxies begin to accrete matter at a rate that increases with time and galactic mergers become more frequent.  This may prevent galaxies from setting into stable configurations, which in turn would cause planetary systems to undergo more frequent close encounters with passing stars.  Life extinctions due to nearby supernova explosions and to gamma-ray bursts would also become more frequent.  Some of these effects have been discussed in Refs.~\cite{Piran:2015yga,Totani:2018zkp}.  With all relevant anthropic effects taken into account, both distributions for $\Lambda$ are likely to be in a good agreement with observation.

\section{Probability distribution for spatial curvature}

In this section we use the 4-volume cutoff measure to calculate the probability distribution for the spatial curvature 
with a cosmological constant fixed at the observed value.
Again, we focus on a subset of bubbles that have the same physical properties as our bubble, 
apart from the e-folding number of the slow-roll inflation inside the bubble, $N_e$.

The spacetime inside a nucleated bubble has a negative spatial curvature. 
After a short period of curvature domination, the curvature rapidly decreases due to inflationary expansion and becomes completely negligible by the end of inflation.  However, it may become significant again in the late universe and may influence structure formation.  
The density parameter for the spatial curvature at present (i.e., at the time when the CMB temperature is the same as in our universe at present), $\Omega_k = 1 - \rho / \rho_{\rm cr}$, where $\rho_{\rm cr}$ is the critical density, 
is related to the e-folding number $N_e$ 
as $\Omega_k \propto e^{-2N_e}$.  
The proportionality constant depends on the detailed history of the universe 
after inflation. 
Since the spatial curvature depends on the reference time 
and the notation for the density parameter may be confused with the 4-volume time, 
we use a time-independent variable $k \equiv (\abs{\Omega_k}^3 / \Omega_\Lambda \Omega_m^2)^{1/3}$ in the following calculation.  
For inflation at the GUT scale 
and assuming instantaneous reheating, $k \sim e^{124-2N_e}$~\cite{DeSimone:2009dq}.

Let us define $\Omega_{\rm nuc}$ as the 4-volume time at bubble nucleation. 
It is related to the time of thermalization $\Omega_*$ as 
\beq 
 \Omega_* = \Omega_{\rm nuc} + \int_{\tau_{\rm nuc}}^{\tau_*} a^3 d \tau 
 = \Omega_{\rm nuc} (1+ C e^{3 N_e}), 
\eeq
where $C$ is a constant that is universal for all bubbles. 
We can neglect the factor of $1$ in the parenthesis 
and obtain $d \Omega_* \propto e^{3N_e} d \Omega_{\rm nuc}$.

As we discussed in the previous section, 
the physical volume nucleating in a 4-volume interval $d \Omega_{\rm nuc}$ is proportional to $d \Omega_{\rm nuc}$. 
After thermalization, the number of observers 
is proportional to $e^{3 N_e} F (\tau_c - \Delta \tau)$ 
and hence the distribution is given by 
\beq
 P (k) dk 
 \propto P_{\rm prior} (N_e (k)) d N_e \int_0^{\Omega_c} e^{3 N_e} F (\tau_c - \Delta \tau) d \Omega_{\rm nuc}, 
\label{Pk0}
\eeq
where the prior distribution $P_{\rm prior} (N_e)$ is determined by the landscape.  Generally we expect that long inflation requires fine-tuning, so $P_{\rm prior} (N_e)$ is a decreasing function of $N_e$.  For a random Gaussian landscape one finds \cite{Masoumi:2016eag, Masoumi:2017gmh}
\beq
P_{\rm prior} (N_e) \propto N_e^{-3}.
\label{PNe}
\eeq

Noting that $F = 0$ for $\Omega_{\rm nuc} \in (\Omega_c / C e^{3 N_e} , \Omega_c)$ 
and $d N_e / d k \propto 1/k$, 
we rewrite Eq.~(\ref{Pk0}) as 
\beq 
P(k) 
 \propto k^{-1} P_{\rm prior} (N_e (k)) \int_0^{\Omega_c} F (\tau_c - \Delta \tau) d \Omega_*. 
 \label{Pk}
\eeq
The proportionality constant is determined by the normalization condition, $\int P(k) dk = 1$. 
Although the integral in \eq{Pk} has the same form as \eq{P}, 
the collapsed fraction $F(\tau)$ is different 
because of the effect of the spatial curvature. 
Again, we use the Press-Schechter form \cite{Press:1973iz,Bardeen:1985tr} 
with a linear perturbation theory for the collapsed fraction $F(\tau)$, 
following Ref.~\cite{DeSimone:2009dq}. 
In that paper, the collapsed function is expressed in terms of 
$x \equiv \rho_\Lambda / \rho_m \propto \tilde{a}^3$. 
Then it is convenient to rewrite 
$X \equiv \Omega_* / \Omega_c$ as
\beq
 X^{-1} \propto \int_0^{\tau_c} \tilde{a}^3 d \tau \propto \int_0^{x_c} \frac{dz}{\sqrt{1 + z^{-1} + k z^{-2/3}}}, 
\eeq
where we use $H^2 = H_\Lambda^2 ( 1 + x^{-1} + k x^{-2/3})$ 
and define $x_c$ as the value of $x$ at $\Omega = \Omega_c$. 
We can calculate \eq{Pk} by rewriting the integral in terms of $x$ 
and using the collapsed function given in Ref.~\cite{DeSimone:2009dq}.

\begin{figure}
\centering
  \includegraphics[width=0.75\linewidth]{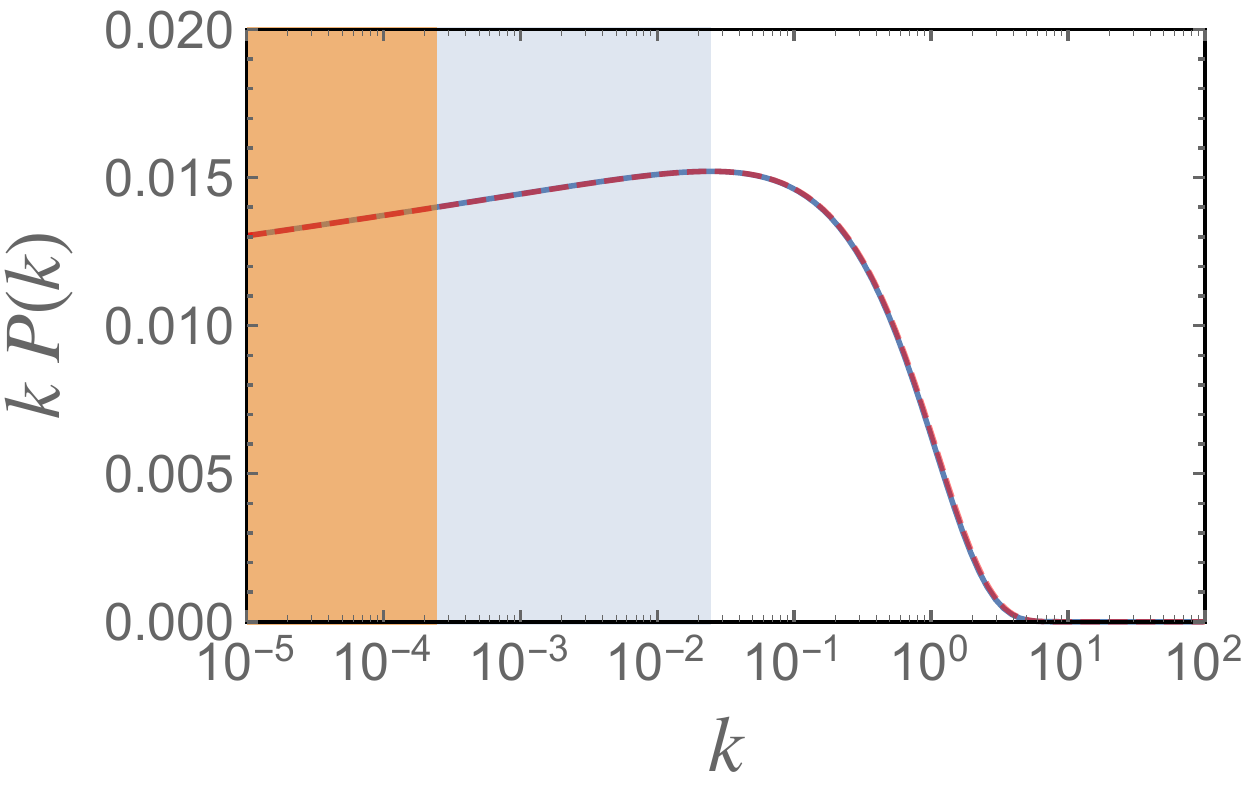}
 \caption{Distribution of spatial curvature in the 4-volume cutoff measure (solid blue curve) and the scale-factor cutoff measure (red dashed curve).  The two distributions are essentially the same.
 The shaded regions are allowed by the Planck constraint. 
 In the blue-shaded region, the spatial curvature may be detected in the future. 
 }
\label{fig3}
\end{figure}

We calculated $P(k)$ numerically with the prior distribution given by Eq.~(\ref{PNe}). 
We neglect $\Delta \tau$ in \eq{Pk} for simplicity because it has been argued in Ref.~\cite{DeSimone:2009dq} that it does not significantly affect 
the collapsed function. 
The result is shown as the solid curve in Fig.~\ref{fig3}. 
This distribution is almost indistinguishable from that in the scale factor cutoff measure~\cite{DeSimone:2009dq} which is shown by a dashed curve. 
The Planck data favors a slightly negative value of $k$~\cite{DiValentino:2019qzk} but is consistent with a spatially flat universe within $2\sigma$~\cite{Aghanim:2018eyx}. The observationally allowed range within $3\sigma$ is about $\abs{\Omega_k} \lesssim 0.01$ or $\abs{k} \lesssim 3 \times 10^{-2}$, which is indicated by shading in the figure. 
The probability for curvature to be in this range is about $94\%$. 
A detection of curvature is probably possible in the future if $k \gtrsim 3 \times 10^{-4}$. 
The range of $k$ where curvature satisfies the observational bound and is still detectable is shown by the blue-shaded region in the figure.
The probability for $k$ to be in this range is about $7\%$ \cite{Freivogel:2005vv,DeSimone:2009dq}.

\section{General formalism}

So far we calculated probability distributions in the 4-volume cutoff measure using the approximate relation (\ref{Omegaa2}) between the scale factor and 4-volume cutoffs.  If a more accurate description is needed, the analysis becomes more complicated.  The reason is that in order to evolve the distribution to larger values of $\Omega$ using 
$d\Omega=a^3 d\tau$, we need to know the scale factor $a$, which generally takes different values on different parts of the constant $\Omega$ surface.  In this section we shall introduce a formalism that can in principle be used to address this issue.

We first consider models where eternal inflation is driven by quantum diffusion of a scalar field $\phi$.  Let us introduce the distribution function $f(\Omega,\phi,V)$ defined as the fraction of comoving volume occupied by regions with given values of $\phi$ and $V=a^3$ on hypersurfaces of constant $\Omega$.  The evolution of the multiverse can then be described by the Fokker-Planck equation \cite{Winitzki:2005ya}
\beq
\frac{\p f}{\p\Omega}+\frac{\p j_\phi}{\p\phi} +\frac{\p j_V}{\p V}= 0,
\label{FP}
\eeq
where the fluxes $j_\phi$ and $j_V$ are given by 
\beq
j_\phi=-\frac{\p}{\p\phi}\left(Df\right)+ \frac{d\phi}{d\Omega} f,
\label{jphi}
\eeq
\beq
j_V = \frac{dV}{d\Omega}f.
\eeq
With $d\Omega=V d\tau$ we can express the drift velocity of $\phi$ as
\beq
\frac{d\phi}{d\Omega}=\frac{1}{V}\frac{d\phi}{d\tau}=-\frac{1}{4\pi V}\frac{dH}{d\phi},
\eeq
where $H(\phi)=[(8\pi/3 U(\phi)]^{1/2}$ is the inflationary expansion rate and $U(\phi)$ is the scalar field potential. 
Similarly, we find
\beq
\frac{dV}{d\Omega}=3{H},
\eeq
where we have used $H=\frac{1}{a}\frac{da}{d\tau}$.

The diffusion coefficient $D$ in Eq.~(\ref{jphi}) can be found from the dispersion of quantum fluctuations of $\phi$ over proper time interval $d\tau$:
\beq
\langle (\delta\phi)^2\rangle = \frac{H^3}{4\pi^2} d\tau = \frac{H^3}{4\pi^2 V} d\Omega = 2D d\Omega, 
\eeq
which gives $D=H^3/8\pi^2 V$.  Combining all this we obtain the following equation for $f(\Omega,\phi,V)$:
\beq
V\left(\frac{\p}{\p\Omega}+3H\frac{\p}{\p V}\right)f-\frac{1}{8\pi^2}\frac{\p^2}{\p\phi^2}(H^3 f) -\frac{1}{4\pi} \frac{\p}{\p\phi} \lmk \frac{d H}{d \phi} f\rmk=0.
\label{diffusion}
\eeq

Once the function $f(\Omega,\phi,V)$ is found, the comoving and physical volume distributions of $\phi$ on surfaces of constant $\Omega$ can respectively be found from
\beq
F(\Omega,\phi)=\int_0^\infty dV f(\Omega,\phi,V)
\label{F}
\eeq
and
\beq
F_V(\Omega,\phi)=\int_0^\infty dV V f(\Omega,\phi,V).
\label{FV}
\eeq

In models with bubble nucleation we can define the distribution $f_j(\Omega,V)$ as the fraction of comoving volume occupied by vacuum of type $j$ with a given value of $V$ on surfaces of constant $\Omega$.  It satisfies the equation
\beq
V\left(\frac{\p}{\p\Omega}+3H_i\frac{\p}{\p V}\right)f_i=\sum_j {\tilde M}_{ij} f_j =\sum_j M_{ij}H_j f_j,
\label{bubble}
\eeq
where ${\tilde M}_{ij}=M_{ij}H_j$ is the proper time transition matrix and $M_{ij}$ is the scale factor time transition matrix given by Eq.~(\ref{Mij}).  The reason we have a proper time transition matrix on the right-hand side of (\ref{bubble}) is that the differential operator $V\p/\p\Omega$ on the left-hand side is 
a derivative with respect to $\tau$.  Once again, the comoving and physical volume distributions of different vacua on surfaces of constant $\Omega$ can be found as
\beq
F_i(\Omega)=\int_0^\infty dV f(\Omega,V)
\label{Fi}
\eeq
and
\beq
F_{iV}(\Omega)=\int_0^\infty dV V f(\Omega,V).
\label{FiV}
\eeq

Equations (\ref{diffusion}) and (\ref{bubble}) are difficult to solve analytically, but they may be useful for a numerical analysis in specific models.  

\section{Summary and discussion}

We have proposed a new probability measure for eternally inflating universes, which regulates infinite numbers of events by a cutoff at a constant 4-volume time $\Omega$, defined by Eqs.~(\ref{Omega0}),(\ref{Omega}). 
The main advantage of this measure is that it avoids the problems with contracting AdS regions that plagued earlier measure proposals.  Otherwise, its properties are similar to those of the scale factor cutoff measure.  With suitable assumptions about the landscape, it does not suffer from the Boltzmann brain problem.  The predicted distribution for the cosmological constant $\Lambda$ is similar to the scale factor measure, but with a higher probability for positive values of $\Lambda$: $P(\Lambda>0)=8\%$ and $3\%$ in 4-volume and scale factor measures, respectively.  
The probability of negative $\Lambda$ is likely to be greatly reduced when anthropic effects in contracting regions are properly taken into account, and one expects the resulting distribution to be in a good agreement with observation.

The probability distribution for the curvature parameter $\Omega_k$ in the new measure is essentially the same as in the scale factor measure, assuming that the cosmological constant is fixed at the observed value.  This distribution depends on the prior distribution $P(N_e)$ for the number of e-foldings of slow roll inflation.  With $P(N_e)\propto N_e^{-3}$, as suggested by random Gaussian models of the landscape, 
one finds that the probability for $\Omega_k$ to be below the observational upper bound ($\Omega_k \lesssim 0.01$) and still be detectable (that is, $\Omega_k \gtrsim 10^{-4}$) is rather small, $P \sim 7\%$.

We note finally that one could introduce a family of measure proposals with properties similar to the 4-volume cutoff.  For example, instead of $\Omega$ one could use the ``time" coordinate
\beq
t_p (\tau)=\int_0^\tau d\tau' [{\cal V}^{(3)}]^p
\eeq
with $p>0$.  The 4-volume cutoff corresponds to $p=1$.  This choice may be preferred because it has a clear geometric meaning.  One hopes however that the probability measure will eventually be determined by the fundamental theory.

\section*{Acknowledgments}
We are grateful to Alan Guth and Ken Olum for useful discussions. 
A.~V. was supported in part by the National Science Foundation under grant PHY-1820872. 
M.~Y. was supported by JSPS Overseas Research Fellowships and the Department of Physics at MIT. 
M.~Y. was also supported by the U.S. Department of Energy, Office of Science, Office of High Energy Physics of U.S. Department of Energy under grant Contract Number DE-SC0012567.

\end{document}